\documentclass[a4paper,11pt]{article}
\usepackage{jcappub}
\usepackage{amsmath}
\usepackage{graphicx}
\usepackage{enumitem}
\usepackage{hyperref}
\usepackage{aas_macros}
\usepackage[]{xcolor}

\makeatletter
\gdef\@fpheader{}
\makeatother

\newlength{\fullw}
\newlength{\halfw}
\newlength{\twofigw}
\newlength{\onefigw}
\DeclareMathOperator{\sign}{sign}

\newcommand{\Mpc}{\mathrm{Mpc}}
\newcommand{\GHz}{\mathrm{GHz}}

\newcommand{\boldmathsymbol}[1]{{\ensuremath{\boldsymbol{#1}}}}
\newcommand{\bm}[1]{\boldmathsymbol{#1}}

\newcommand{\CLASS}{\textsc{class}}
\newcommand{\CAMB}{\textsc{camb}}
\newcommand{\MP}{\textsc{MontePython}}
\newcommand{\COSMOMC}{\textsc{CosmoMC}}

\newcommand{\calC}{\mathcal{C}}

\newcommand{\dd}{{\mathrm{d}}}
\newcommand{\uini}{\mathrm{ini}}
\newcommand{\ub}{\mathrm{b}}
\newcommand{\ue}{\mathrm{e}}

\newcommand{\us}{\mathrm{s}}
\newcommand{\ui}{\mathrm{i}}
\newcommand{\ureio}{\mathrm{reio}}
\newcommand{\xe}{x_\ue}
\newcommand{\xet}{\tilde{x}_\ue}

\newcommand{\uL}{\mathrm{L}}

\newcommand{\blur}{\mathrm{blur}}
\newcommand{\lens}{\mathrm{lens}}
\newcommand{\blurlens}{\blur-\lens}

\newcommand{\etaini}{\eta_{\uini}}
\newcommand{\Ai}{A_{\ui}}

\newcommand{\alphai}{\alpha_{\ui}}
\newcommand{\Ri}{R_{\ui}}
\newcommand{\gammai}{\gamma_{\ui}}
\newcommand{\An}{A}

\newcommand{\alphan}{\alpha}
\newcommand{\Rn}{R}
\newcommand{\gamman}{\gamma}
\newcommand{\Ax}{A_{\times}}

\newcommand{\alphax}{\alpha_{\times}}
\newcommand{\Rx}{R_{\times}}

\newcommand{\fitconst}[3]{#1_{#2_{#3}}}
\newcommand{\aione}{\fitconst{a}{\ui}{1}}
\newcommand{\aitwo}{\fitconst{a}{\ui}{2}}
\newcommand{\Rione}{\fitconst{R}{\ui}{1}}
\newcommand{\Ritwo}{\fitconst{R}{\ui}{2}}
\newcommand{\Rithree}{\fitconst{R}{\ui}{3}}
\newcommand{\alpizero}{\fitconst{\alpha}{\ui}{0}}
\newcommand{\alpione}{\fitconst{\alpha}{\ui}{1}}
\newcommand{\gamizero}{\fitconst{\gamma}{\ui}{0}}
\newcommand{\gamione}{\fitconst{\gamma}{\ui}{1}}
\newcommand{\gamitwo}{\fitconst{\gamma}{\ui}{2}}

\newcommand{\axone}{\fitconst{a}{\times}{1}}
\newcommand{\axtwo}{\fitconst{a}{\times}{2}}
\newcommand{\Rxzero}{\fitconst{R}{\times}{0}}
\newcommand{\Rxone}{\fitconst{R}{\times}{1}}
\newcommand{\Rxtwo}{\fitconst{R}{\times}{2}}
\newcommand{\alpxzero}{\fitconst{\alpha}{\times}{0}}
\newcommand{\alpxone}{\fitconst{\alpha}{\times}{1}}
\newcommand{\alpxtwo}{\fitconst{\alpha}{\times}{2}}

\newcommand{\anone}{\fitconst{a}{1}{}}
\newcommand{\antwo}{\fitconst{a}{2}{}}
\newcommand{\Rnone}{\fitconst{R}{1}{}}
\newcommand{\Rntwo}{\fitconst{R}{2}{}}
\newcommand{\Rnthree}{\fitconst{R}{3}{}}
\newcommand{\alpnzero}{\fitconst{\alpha}{0}{}}
\newcommand{\alpnone}{\fitconst{\alpha}{1}{}}
\newcommand{\gamnzero}{\fitconst{\gamma}{0}{}}
\newcommand{\gamnone}{\fitconst{\gamma}{1}{}}
\newcommand{\gamntwo}{\fitconst{\gamma}{2}{}}

\newcommand{\zreio}{z_\ureio}

\title{Lensing anomalies from the epoch of reionisation}

\author[a]{Christian Fidler,}
\author[a]{Julien Lesgourgues}
\author[b]{and Christophe Ringeval}

\affiliation[a]{Institute for Theoretical Particle Physics and
  Cosmology (TTK), RWTH Aachen University, D-52056 Aachen, Germany}

\affiliation[b]{Cosmology, Universe and Relativity at Louvain,
  Institute of Mathematics and Physics, Louvain University, 2 Chemin
  du Cyclotron, 1348 Louvain-la-Neuve, Belgium}

\emailAdd{julien.lesgourgues@physik.rwth-aachen.de}
\emailAdd{fidler@physik.rwth-aachen.de}
\emailAdd{christophe.ringeval@uclouvain.be}

\date{today}

\begin{document}

\abstract{Reionisation blurring is a non-linear correction to the
  cosmic microwave background that acts similar to weak gravitational
  lensing and that can be computed from linear perturbations through a
  blurring potential. Its impact on the cosmic microwave background is
  roughly two order of magnitude smaller than that of lensing, in
  isolation. But the blurring potential is strongly correlated with
  the lensing potential thereby generating a potentially observable
  cross-correlation. We compute for the first time the inclusive
  impact of reionisation blurring on the temperature angular power
  spectrum and discuss how much it could induce lensing anomalies.}


\maketitle

\section{Introduction}

\label{sec:intro}
The cosmic microwave background has shaped our understanding of the
Universe over the past decades and remains of central importance for
precision cosmology. Parameter estimation relies crucially on the
Planck data~\cite{Aghanim:2018eyx} and thus the accurate analysis of
secondary effects that may contaminate the cosmic microwave background
(CMB) is an important task. Secondary CMB anisotropies are not only
foregrounds for cosmological data analysis, they also carry additional
and unique information on the late time universe and the non-linear
growth of cosmic structures. The Planck data are accurate enough to be
sensitive to various of these secondary anisotropies. The
non-Gaussianities contained in the CMB have been measured for the
first time in Ref.~\cite{Ade:2015ava}, they are primarily induced by
weak gravitational lensing which describes the bending of light by the
gravitational potential of the matter along the
line-of-sight~\cite{Bartelmann:1999yn,Lewis:2006fu}. Planck is also
sensitive to the Cosmic Infrared Background (CIB) and thousands of
galaxy clusters have been detected through localised
Sunyaev-Zel'dovich (SZ) signals in the CMB
maps~\cite{2016A&A...594A..23P, 2016A&A...594A..24P}. Some
cosmological parameters can also be extracted from SZ cluster counts
and lensing measurements, alone, and this allows for an intricate
testing of the $\Lambda$CDM model over the whole cosmological
history. Although the lensing data are in agreement with CMB primary
anisotropies on a wide range of length scales, a three-sigma deficit
in ``curl power'' has been consistently reported within the multipole
range $\ell \in [264,901]$~\cite{2016A&A...594A..15P,
  Aghanim:2018oex}.

The reionisation of the universe by the first stars is another source
of unavoidable secondary CMB anisotropies which is usually accounted
for in data analysis through an unique homogeneous parameter, the
reionisation optical depth. It describes the suppression of
perturbations and, although strongly correlated with the primordial
power spectrum amplitude, its value can be accurately inferred from
the $E$-mode polarisation induced by the free electrons created during
the Epoch of Reionisation (EoR)~\cite{2016A&A...596A.108P}. By
definition, reionisation is a highly non-linear process in which the
physics of very non-linear structures (stars) backreacts onto the
largest length scales of the universe. One may therefore question the
accuracy of modelling reionisation effects on the CMB by only the
homogeneous optical depth~\cite{Santos:2003jb}, while in reality
reionisation is always a patchy process. In
Ref.~\cite{Fidler:2017irr}, we have shown that anisotropies in the
reionisation optical depth induces various new secondary distortions
on the CMB anisotropies. They are induced by inhomogeneities in the
baryon density, in the ionisation fraction, in gravitational redshift,
while some of these effects can be identified as the diffuse analogues
of blurring and kinetic SZ effects~\cite{Sunyaev:1970, Sunyaev:1980,
  HernandezMonteagudo:2009ma, Yasini:2016pby}. Because inhomogeneities
in the ionisation fraction trace cosmic structures, one may wonder if
it could not be the source of the systematic lensing deviations
currently measured in the CMB data. Among all these effects, the
diffuse blurring at reionisation could be the main suspect. It is a
loss term, i.e., it describes an incoming bundle of light rays which
is scattered out of the observer line-of-sight due to, from some
parts, collisions with ionised gas created by the ionising radiation
of the first stars. As we discuss below, it is expected to be
correlated with lensing and this correlation has never been estimated
before. Let us stress that the impact of a ionised clouds has been
well studied in the literature \cite{Dvorkin:2008tf,
  2009PhRvD..79j7302D, 2013ApJ...776...82N}, focusing on the so-called
``screening'' contribution related to the ionisation fraction and
density of the absorbing cloud. This is the main driver of the
blurring effect, however, in this analysis, we consider two additional
terms that have not been studied before; a relativistic and a kinetic
correction that naturally appear from our inclusive second-order
Boltzmann analysis. These terms are particularly relevant on
intermediate scales, around $l=100$~\cite{Fidler:2017irr}.

Mathematically lensing can be described as a modulation of the primary
CMB by the deflections that the gravitational potentials of the
forming overdensities induce. Lensing is implemented in the most
popular Boltzmann codes \cite{Lewis:1999bs, Blas:2011rf}, employing
the decomposition of the non-linear effect into a convolution of two
independent linear terms, the primary CMB and the lensing potential
\cite{Lewis:2006fu}. Let us notice that lensing is not the only
modulation of the microwave background along the
line-of-sight. Gravitational redshift modulations describe the impact
of an inhomogeneous expansion, and time-delay effects change the
distance to the last-scattering surface. But both are significantly
smaller than lensing and cannot explain the observed missing
power~\cite{Mollerach:1997up, Hu:2001yq, Fidler:2014zwa}.

While the physical origin of lensing and reionisation blurring are
different, the way both effects impact the CMB is similar. Lensing
describes how the primary CMB is distorted by the late Universe
gravitational potentials and blurring is the distortion accompanying
ionised clouds. As for the lensing, we can implement the blurring
effect by defining a \emph{blurring} potential, that is sourced at
late times and correlates with the matter
overdensities~\cite{Fidler:2017irr}. Beyond linear perturbations, the
probability to be scattered depends on the distribution of matter
along the path of the CMB photons such that the overall effect is more
than just dimming the amplitude, the angular power spectrum is
distorted, or \emph{blurred}. In Ref.~\cite{Fidler:2017irr}, we have
shown that this blurring contribution is the dominant second-order
effect along the line-of-sight after lensing. It provides corrections
that are typically two orders of magnitude smaller, a priori not large
enough to be visible by Planck. But lensing and blurring may be
correlated, generating a cross-term that could potentially cause much
larger corrections (see Ref.~\cite{Feng:2018eal} for an analysis based
on the screening terms only).

In this paper, we derive and compute the non-linear blurring-lensing
correlation for the intensity $I$, and we estimate whether it can lead
to significant corrections to the lensing signal. The paper is
organised as follows. In section~\ref{sec:kappablur} we summarise our
notations, review the blurring effect, and describe our method to
compute the cross-correlation with lensing. In section~\ref{sec:res}
we discuss our numerical results and perform a statistical analysis
using the latest public Planck likelihoods and data, which currently
are the 2015 ones~\cite{Ade:2015xua}. Finally, we discuss our results
and conclude.

\section{Reionisation blurring}
\label{sec:kappablur}

\subsection{Notation and conventions}

Our convention follows the one presented in Refs.~\cite{Beneke:2010eg,
  Beneke:2011kc, Fidler:2017irr} that we briefly recap below. The
metric is assumed to be of the form
\begin{equation}
\dd s^2 = a^2\left\{(1+2A)\dd \eta^2 + 2 B_i \dd\eta \dd x^i -
\left[(1+2D)\delta_{ij} + 2 E_{ij}\right]\dd x^i \dd x^j\right\},
\end{equation}
where $A$ is the lapse perturbation, $\bm{B}\equiv\{B^{i}\}$ the shift
vector, $D$ stands for the spatial trace perturbation and $E_{ij}$ is
the symmetric spatial stress tensor. We use second-order perturbation
theory in which the stress tensor, the Einstein equation and the
Boltzmann equation are all expanded as $X = X^{(1)} + X^{(2)}
+\cdots$. Simplifications are made using $B_i^{(1)}= E^{(1)}_{ij} =
0$, corresponding to the so-called Poisson gauge at linear order. We
will further assume that all non-scalar modes are of second or higher
order.

The stress-energy tensor is decomposed as
\begin{equation}
T_{\mu\nu} = (\rho+P)u_\mu u_\nu - P g_{\mu\nu} + \Sigma_{\mu\nu},
\end{equation}
where $\rho$ is the energy density, $P$ the pressure density,
$\Sigma_{\mu\nu}$ the anisotropic stress tensor and $u_\mu$ the
rest-frame 4-velocity.
 
Cold species, such as massive particles, are uniquely described by
their energy density $\rho$ and their 3-velocity $\bm{v}$. The latter is
defined as the spatial part of $u$ at linear order, $v^{(1)}_i \equiv a
u^{(1)}_i$. On the contrary, the complex phase-space of relativistic species
requires to specify the full distribution function $f(\eta,
x, q \bm{n})$ where the comoving momentum $\bm{q}=a \bm{p}$
has been expressed as a direction $\bm{n}$ and a magnitude $q$.  We define the
integrated distribution
\begin{equation}
\Delta(\eta,\bm{x},\bm{n}) \equiv \dfrac{\int \dd q q^3
  f(\eta, x, q \bm{n})}{\int \dd q q^3 f^{(0)}(q)}\,,
\end{equation}
that can be understood as a temperature perturbation and which
contains all the information required to evaluate the stress-energy
tensor of a relativistic specie. We use the Fourier
conventions
\begin{equation}
  A(\bm{x}) = \int \frac{\dd^3 \bm{k}} {(2\pi)^3} e^{i\bm{k} \cdot \bm{x}}
  A(\bm{k})\,,
\end{equation}
Finally, we use a shortcut notation for convolutions which appear as
multiplications of different wavenumbers, namely
\begin{equation}
A(\bm{k}_1) \cdot B (\bm{k}_2)  \equiv \int \dfrac{\dd^3\bm{k}_1}{(2\pi)^3}
\int \dfrac{\dd^3\bm{k}_2}{(2\pi)^3}(2\pi)^3\delta^3(\bm{k}-\bm{k}_1
-\bm{k}_2)A(\bm{k}_1)B(\bm{k}_2)\,.
\end{equation}

\subsection{Blurring, lensing and their correlation}

The differential equation for non-linear blurring of the intensity
perturbation to second order in perturbation theory, $\Delta^{(2)}$,
has been derived in Ref.~\cite{Fidler:2017irr} and reads, in Fourier
space,
\begin{equation}
\dot{\Delta}^{(2)} + i \bm{n} \cdot \bm{k}
\Delta^{(2)} = -|\dot{\kappa}|\Delta^{(2)} 
- |\dot{\kappa}| \left(A^{(1)} +
\delta_\ub^{(1)} + \delta^{(1)}_{\xe} -\bm{n} \cdot \bm{v}^{(1)}_\ub
\right)(\bm{k}_1) \cdot \Delta^{(1)}(\bm{k}_2)\,,
\label{eq:boltzloss}
\end{equation}
where $\delta_\ub = \delta\rho_\ub / \bar\rho_\ub$ is the relative
perturbation to the baryon density, and $\delta_{\xe}$ that to the
ionisation fraction. The left-hand side describes the free propagation
of photons in the direction $\bm{n}$ while the right-hand side
represents collisions scattering photons out of the observer
line-of-sight. The first term is ``purely'' second-order. It features
an isotropic collision rate $|\dot\kappa|$, and it leads to an overall
suppression similar in all points to the linear theory. The following
terms describe an anisotropic modulation of the reaction rate. They
enhance the scattering probability in overdense regions,
$\delta_\ub^{(1)}>0$, and in ionised patches,
$\delta^{(1)}_{\xe}>0$. These are commonly referred to as
``screening'' while the term in $\delta_\ub^{(1)}$ is the diffuse
analog of the so-called blurring SZ
effect~\cite{HernandezMonteagudo:2009ma, Fidler:2017irr}. There are
further a kinematic enhancement based on the motion of the ionised gas
with respect to the CMB rest frame, $\bm{n} \cdot \bm{v}^{(1)}_\ub$
(blurring Doppler term), plus a relativistic correction $A^{(1)}$ due
to shift between the cosmological time $\eta$ and the time coordinate
of the inertial frame comoving with the baryons (blurring GR term).

As explained in Ref.~\cite{Fidler:2017irr}, we may integrate the
probability for collisions along the line-of-sight. This allows us
to define the \emph{blurring potential}:
\begin{equation}
\begin{aligned}
  \kappa^{(1)}_{\blur}(\eta,\bm{k}_1,\bm{n}) & \equiv
  -\int_{\etaini}^{\eta} \dd \eta' e^{- i \bm{n} \cdot \bm{k}_1 (\eta
    - \eta')} \left|\dot{\kappa}(\eta') \right| \\ & \times
  \left[\bm{n} \cdot \bm{v}^{(1)}_\ub(\eta',\bm{k}_1) -
    A^{(1)}(\eta',\bm{k}_1) - \delta^{(1)}_\ub(\eta',\bm{k}_1) -
    \delta^{(1)}_{\xe}(\eta',\bm{k}_1) \right],
\end{aligned}
\end{equation}
with the help of which equation~\eqref{eq:boltzloss} can be integrated as
\begin{equation}
\Delta^{(2)}(\eta,k) = - \kappa^{(1)}_{\rm blur}(\bm{k}_1) \cdot
\Delta^{(1)}(\bm{k}_2)\,.
\end{equation}
In real space, we then find a simple equation separating the
second-order terms into a product of linear perturbations at the
present time
\begin{equation}
\Delta^{(2)}(\eta,\bm{x},\bm{n}) =
-\kappa^{(1)}_{\blur}(\eta,\bm{x},\bm{n})
\Delta^{(1)}(\eta,\bm{x},\bm{n}).
\label{eq:2ndblur}
\end{equation}
This expression shows that the blurring potential mathematically
appears as an inhomogeneous correction to the usual (homogeneous)
optical depth $\kappa$. As such, at second order, it can be
conveniently included in the line of sight integrations by
exponentiation
\begin{equation}
\begin{aligned}
  \Delta(\eta,\bm{x},\bm{n}) & =
\exp\left[-\kappa(\eta) -\kappa^{(1)}_{\blur}(\eta,\bm{x},\bm{n}) \right]
\Delta^{(1)}(\eta,\bm{x},\bm{n}) \\ & \simeq
e^{-\kappa(\eta)} \left[ \Delta^{(1)}(\eta,\bm{x},\bm{n}) -
\kappa^{(1)}_{\blur}(\eta,\bm{x},\bm{n})
\Delta^{(1)}(\eta,\bm{x},\bm{n})\right] .
\end{aligned}
\label{eq:expblur}
\end{equation}
where the first term gives back the linear result and the second term
equation~\eqref{eq:2ndblur}. Defining the blurring potential as above
implicitly requires that we neglect any reionisation blurring on the
late-time Integrated Sachs-Wolf (ISW) perturbations. This additional effect may lead to
small corrections on the largest angular scales, but the same
approximation is routinely applied in the standard lensing computation
and is necessary to be able to separate lensing into a product of two
linear terms without an intrinsic second-order contribution. Our
equation indeed shows a remarkable similarity to lensing, where the angular
power spectrum can also be written as a modulation of the linear
result \cite{Lewis:2006fu}:
\begin{equation}
\Delta^{(2)}_{\lens}(\eta,\bm{x},\bm{n}) = \nabla^a
\Psi^{(1)}_{\lens}(\eta,\bm{x},\bm{n}) \nabla_a
\Delta^{(1)}(\eta,\bm{x},\bm{n}) ,
\end{equation}
with $\nabla_a$ the angular derivative .

Next we employ the flat sky limit, accurate on scales $l>10$. In this
limit, the curvature of the celestial sphere is neglected, simplifying
the angular multipole decomposition into a Fourier analysis. For the
blurring, the flat sky limits gives
\begin{equation}
\Delta_{\blur,I}^{(2)}(l) = - \int \frac {\dd^2 l'}{(2\pi)^2}
\kappa^{(1)}_{\blur}(l-l') \Delta^{(1)}_{I}(l')\,,
\end{equation}
where $\bm{l}$ is a 2-vector parametrising the flat sky, that would be
replaced by a discrete set of multipole coefficients $(l,m)$ on the
sphere.

A reasonable hypothesis is to assume that the correlation between
$\kappa_{\blur}$ and $\Delta^{(1)}$ is suppressed since the linear
perturbations arise from the time of recombination in the early
Universe while the blurring potential is sourced only in the late
Universe. In the flat sky limit, the angular power spectra for the
intensity is then given by
\begin{equation}
\calC_{\blur,I}(l) = \int \dfrac{\dd^2 l'}{(2\pi)^2}
\calC^{\kappa}(l-l') \calC_{I}(l') - \calC_{I}(l) \int \dfrac {\dd^2
  l'}{(2\pi)^2} \calC^{\kappa}(l'),
  \label{eq:blur}
\end{equation}
where the last term is obtained by expanding equation~\eqref{eq:expblur} up
to second order and with the present day angular temperature power
spectrum $ 16\delta(l-l')\calC_I(l) = \langle \Delta(l) \Delta(l')^{*}
\rangle$ and the present day spectra of the blurring and lensing
potentials, $\delta(l-l')\calC^{\Psi}(l) = \langle \Psi(l)
\Psi^{*}(l') \rangle$, $\delta(l-l')\calC^{\kappa}(l) = \langle
\kappa^{\phantom{*}}_{\blur}(l)\kappa^*_{\blur}(l') \rangle$ and
$\delta(l-l')\calC^{\kappa\Psi}(l) = \langle \kappa_{\blur}(l)
\Psi^{*}(l') \rangle$.  This allows us to compare this expression with
the one used to compute the lensing, namely~\cite{Lewis:2006fu}
\begin{equation}
\calC_{\lens,I}(l) = \int \frac {\dd^2 l'}{(2\pi)^2} \left[\bm{l}'
  \cdot (\bm{l}-\bm{l}') \right]^2 \calC^{\Psi}(l-l') \calC_{I}(l') -
\calC_{I}(l) \int \dfrac {\dd^2 l'}{(2\pi)^2} (\bm{l} \cdot \bm{l}')^2
\calC^{\Psi}(l').
\label{eq:lens}
\end{equation}
The only differences are exchanging the blurring potential for the
lensing one and the appearance of additional factors $\bm{l}$ due to
the angular derivatives.

The correlation between lensing and blurring can be derived in a
similar manner and one gets
\begin{equation}
\calC_{\blurlens,I}(l) = \int \dfrac {\dd^2 l'}{(2\pi)^2}
\left[\bm{l}' \cdot (\bm{l}-\bm{l}')\right] \calC^{\kappa\Psi}(l-l')
\calC_{I}(l') - \calC_{I}(l) \int \dfrac {\dd^2 l'}{(2\pi)^2} (\bm{l}
\cdot \bm{l}') \calC^{\kappa \Psi}(l').
\label{eq:blur-lens}
\end{equation}
Notice that the second term vanishes due to the antisymmetry of the
integrand in $\bm{l}'$.

The blurring potential can be computed using a linear Boltzmann code
following the implementation of the lensing potential. In the flat sky
limit the angular power spectrum can then be constructed via the above
integrals. We have implemented the blurring computation using our
equations in the Boltzmann code {\CLASS}~\cite{2011arXiv1104.2932L,
  Blas:2011rf}, in the Newtonian gauge.

\subsection{Epoch of Reionisation}
\label{sec:eor}

\begin{figure}
  \begin{center}
    \includegraphics[width=\onefigw]{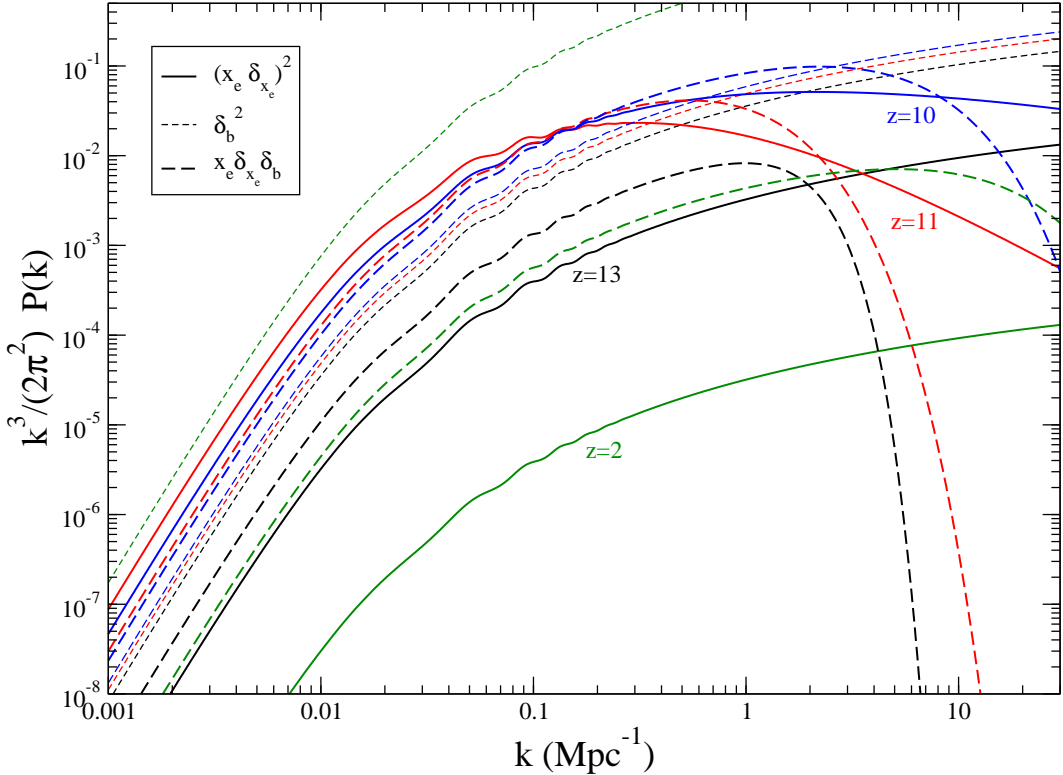}
    \caption{Typical evolution of the ionizing power spectra during
      and after the EoR as a function of redshift. Solid curves are
      for the ionized fraction, long dash curves for the
      cross-correlation between ionized fraction and baryons, and
      short dash curves are the power spectrum of baryons. The
      different colours show these spectra at redshift $z=13$, $z=11$,
      $z=10$ and $z=2$ for a fiducial reionisation model at $\zreio
      \simeq 12$.}
    \label{fig:reiopk}
  \end{center}
\end{figure}

While most of the sources for the blurring potential can be directly
extracted from the linear perturbations computed in {\CLASS}, the
perturbation of the ionisation fraction $\delta_{\xe}$ requires a
different framework. We have used a similar approach as in
Refs.~\cite{Mao:2008ug, Clesse:2012th, Fidler:2017irr}, namely, using
analytical expressions fitted against numerical simulation
results~\cite{McQuinn:2007dy, Zahn:2006sg, Jelic:2008jg, Zahn:2010yw,
  Iliev:2015aia, Lin:2015bcw, Bauer:2015tta}. More precisely, the
functional form of the relevant power spectra is the same as in
Ref.~\cite{Mao:2008ug} and given by
\begin{align}
  \xet^2 P_{\delta_{\xe}\delta_{\xe}}(\eta,k) & = \Ai^2(\xet) \left(1 -
  \xet \right)^2 \left\{1 + \alphai(\xet) k \Ri(\xet) + \left[k
    \Ri(\xet) \right]^2 \right\}^{-\gammai(\xet)/2}
  P_{\delta_{\ub}}(\eta,k),\label{eq:reio_fit} \\
  \xet P_{\delta_\ub \delta_{\xe}}(\eta,k) & = \Ax(\xet) \left(1 -
  \xet \right) \exp\left\{-\alphax(\xet) k \Rx(\xet) - \left[k
    \Rx(\xet) \right]^2\right\} P_{\delta_{\ub}}(\eta,k),
\label{eq:reio_cross}
\end{align}
where $\xet(\eta)\equiv \xe(\eta)/\xe(\eta_0)$ is the background
normalised ionised fraction, and $\xe(\eta_0)$ is the total ionised
fraction once the Universe is completely reionised. The various
functions of $\xet$ entering equations~\eqref{eq:reio_fit} and
\eqref{eq:reio_cross} encode various physical evolving properties. For
instance the length scale $R(\xe)$ keeps track of the typical ionised
bubble size while $P_{\delta_{\ub}}(\eta,k)$ is the baryon power
spectrum whose evolution is solved in {\CLASS}. The fitting form, with
respect to $\xet$, of all the functions appearing in
equations~\eqref{eq:reio_cross} are detailed in
Appendix~\ref{appendix}.

In figure~\ref{fig:reiopk}, we have represented the typical evolution
of these spectra for a given reionisation model with an optical depth
$\tau_\ureio=0.0952$ and a reionisation redshift $\zreio \simeq 12$.
This figure shows that, up to different relative amplitudes, the
spectrum of the ionisation fraction as well as its correlation with
the baryons, and the power spectrum of baryons all evolve in parallel
up to some relatively small scales $k \gtrsim 0.2$ (from the CMB point
of view). For this reason, we have adopted a much simplified model
and, in {\CLASS}, we choose directly $\delta_{\xe}$ to be of a
functional form given by the square root of its power spectrum, namely
\begin{equation}
  \delta_{\xe}(\eta,k) = \An(\xet) \left(1 - \xet \right) \left\{1 +
  \alphan(\xet) k \Rn(\xet) + \left[k \Rn\xet) \right]^2
  \right\}^{-\gamman(\xet)/4} \delta_{\ub}(\eta,k)\,.
  \label{eq:reio_sqrt}
\end{equation}
The unknown functions in this equation, $\An(\xet)$, $\alphan(\xet)$,
$\Rn(\xet)$ and $\gamman(\xet)$ should be viewed as nuisance. They
have been chosen of the same functional form as their $\Ai(\xet)$,
$\alphai(\xet)$, $\Ri(\xet)$ and $\gammai(\xet)$ counterparts but
instead of being fitted to reionisation simulation data, we let all
their parameters free to vary over some conservative prior range
encompassing simulation results. In total, we end up with ten new free
parameters. More details on the modelling are given in
appendix~\ref{appendix}. Let us stress again that our goal here is not
to have the most accurate reionisation sources model but rather to
marginalise over the nuisances they are expected to induce, see
section~\ref{sec:cmb}.

\section{Lensing anomalies}
\label{sec:res}
By computing the level of correlation between blurring and lensing, we
are in the position to calculate the total impact of blurring on the
angular power spectrum. Including lensing, but omitting blurring in
theoretical predictions, is therefore expected to artificially induce
deviations between the measured power spectrum and the predictions,
thereby potentially triggering an artificial ``lensing anomaly''. In
this section, we calculate the amplitude of such an effect onto the
power spectrum.

\subsection{Lensing and blurring potentials}

The correlation between the blurring and lensing potential comes from
the fact that both effects are induced by the late Universe matter
overdensities. These overdensities source the gravitational
potentials, which are responsible for lensing, and are the seeds of
reionisation.

In order to solve equation~\eqref{eq:blur-lens}, one needs the correlator
$\calC^{\kappa \Psi}$. Both potentials, $\kappa_{\blur}$ and $\phi$,
can be computed from the linear perturbations, and thus by using
{\CLASS}. As discussed before, the only caveat in the calculation of
blurring is the use of a analytical fit for the perturbed ionisation
fraction, see equations~\eqref{eq:reio_sqrt}. Taking the best fit values
mentioned in the appendix~\ref{appendix}, our results for the spectrum
of the two potentials and their cross-correlation are shown in
figure~\ref{fig:potential}. By comparing the three curves, we see that
blurring and lensing are correlated on all but the largest
scales. This is expected because, on very large scales, reionisation
is no longer tracing overdensities, but becomes dominated by the
expansion of the ionised bubbles instead. Note that these scales are
particularly relevant for lensing.

\subsection{Lensing-blurring corrections to the angular power spectrum}

\begin{figure}
\begin{center}
\includegraphics[width=\onefigw]{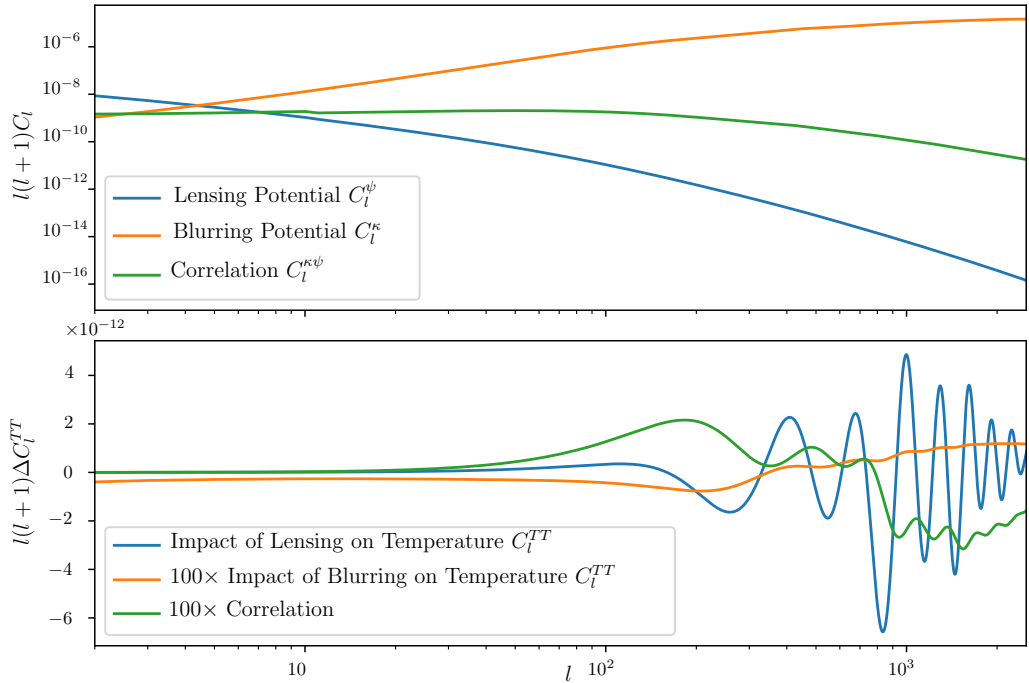} 
\caption{In the upper panel, the lensing potential (lower blue curve)
  dominates on the largest scales, while the blurring potential (top
  orange curve) stays relevant up to much smaller scales. Their
  correlation is shown in green (middle curve), and remains in-between
  both potentials on all but the largest scales. This implies a strong
  correlation for most multipoles. Note that the relative amplitude of
  the lensing and blurring potentials is not representative of their
  respective impact on the angular power spectrum (shown in the lower
  panel), since a different number of angular derivatives are present
  in equations~\eqref{eq:blur} to \eqref{eq:blur-lens}.  The impact on
  the angular temperature power spectra is shown in the lower plot,
  where blurring and the correlation between blurring and lensing is
  boosted by a factor of 100 for better visibility.  The impact of
  blurring is relatively smooth because the blurring potential has
  support up to the much smaller scales. As such, the convolution
  integral of equation~\eqref{eq:blur} runs over many acoustic
  oscillations of the primary CMB and washes them out. The correlation
  between lensing and blurring is of the same order of magnitude as
  blurring, and shows oscillations with the same frequency as lensing,
  but with a $\pi/2$ shift in phase.}
\label{fig:potential}
\end{center}
\end{figure}

Having computed the auto- and cross-correlation spectra of the the
blurring and lensing potentials, we employ equations~\eqref{eq:blur} to
\eqref{eq:blur-lens} to obtain the final impact of lensing and
blurring on the temperature angular power spectrum, shown in the lower
panel of figure~\ref{fig:potential}, where the effect of blurring and
of its correlation to lensing have been multiplied by a factor of
$100$ for better visibility.  We notice that the correlation
deceptively ends up being of similar magnitude as the blurring
effect. Looking at the potential spectra in the upper panel of
figure~\ref{fig:potential}, one could have expected the correlation
effect to be much larger than the blurring one. This ends up not being
the case due to the absence of correlation between the lensing and
blurring potentials on the largest scales, which are responsible for a
significant part of the lensing signal.

The correlation also shows oscillations comparable to the lensing
signal. In principle, when fitting the observational data (which are
affected by all these effects) by a theoretical model in which
blurring is neglected, one would therefore expect to get a small
mismatch, that could be misinterpreted as an excess, or a deficit, of
lensing. This could be the case especially if the oscillations in the
lensing and in the lensing-blurring correlation signal had the same
phase.  However, the lower panel of figure~\ref{fig:potential} shows
that the lensing-blurring oscillations are shifted in phase, roughly
by $\pi/2$. It is thus less clear whether the lensing-blurring effect
can be confused with a lensing effect of different magnitude, or if
this shift could actually be preferred by the data. Furthermore the
amplitude of these corrections is small, typically around $1\%$.

\section{CMB data analysis with lensing-blurring}
\label{sec:cmb}

Because reionisation blurring is a non-speculative contribution that
is present in the CMB, we now address the question whether neglecting
it may bias the extracted cosmological parameters, or could be
misinterpreted as some kind of anomaly in the data.

For this purpose, we fitted the Planck 2015 TT+lowP
data~\cite{Ade:2015xua} to different models and sets of free
parameters, using our modified version of
{\CLASS}\footnote{\url{http://class-code.net}}
v2.6.3~\cite{Blas:2011rf} in combination with the parameter extraction
code
{\MP}\footnote{\url{https://github.com/brinckmann/montepython_public}}
v3.1.0~\cite{Audren:2012wb,Brinckmann:2018cvx}.  As discussed in
section~\ref{sec:eor}, the details of the reionisation which affect
the blurring are encoded in ten nuisance parameters that we
incorporate in the data analysis. This allows us to check if CMB
observations prefer some inhomogeneous reionisation evolution. Note
that we do not include the Planck CMB lensing likelihood in our
analysis. This likelihood accounts for the extraction of the lensing
signal from the trispectrum of observed temperature and polarization
anisotropies. In principle, blurring effects also contribute to this
trispectrum, but the standard lensing extraction process neglects this
contribution. It is beyond the scope of this paper to compute whether
the blurring effect has a significant impact on the lensing extraction
pipeline or not, and whether the CMB lensing estimator should be
corrected accordingly. Thus, we conservatively ignore lensing
extraction, and focus only on the temperature (and low-l polarization)
power spectra measured by Planck. See Refs.~\cite{2013PhRvD..87d7303G,
  Namikawa:2017uke} for searches on the trispectrum that would be
induced by the screening terms only.

\subsection{Bias on cosmological parameters}

To check the impact of neglecting blurring on the measurement of
standard cosmological parameters, we first fitted the data with the
exact same baseline $\Lambda$CDM model as in \cite{Ade:2015xua} (with
flat priors on the six standard parameters $\{ \omega_\mathrm{b},
\omega_\mathrm{cdm}, \theta_\mathrm{s}, \ln( 10^{10} A_\mathrm{s}),
n_\mathrm{s}$, $\tau_\mathrm{reio}\}$ and standard priors on the
Planck nuisance parameters), switching off the blurring effect as in
all previous analyses. We checked the level of agreement between runs
performed with {\MP}+{\CLASS} in the synchronous gauge, in the
newtonian gauge (that we will always use when switching on blurring),
and with {\COSMOMC}+{\CAMB}~\cite{Lewis:2002ah,Lewis:1999bs} in the
synchronous gauge. For the latter we use the results publicly
available on the Planck Legacy
Archive\footnote{\url{https://pla.esac.esa.int}}, quoted in subsection
2.1 of the PDF parameter table document. All three cases are based on
8 Markov chains that have reached a Gelmann-Rubin convergence
criterium $R-1 \leq 0.01$ for all parameters. We find that between the
two runs in the synchronous gauge, error bars always agree with each
other within $5\%$, and mean values up to $\pm 0.09 \sigma$. Between
the {\CLASS} runs in the two different gauges, error bars agree within
$4\%$, and mean values up to $\pm 0.09 \sigma$ again. We conclude that
in general, with standard precision settings and convergence criteria,
results can be trusted up to approximately $\pm 0.1 \sigma$ for the
mean and $5\%$ for the error. This finite precision comes from: the
level of convergence of the MCMC chains; numerical errors when
deriving confidence limits from binned chains; and the precision of
{\CAMB} and \CLASS{} when they are used with default
precision\footnote{With very high precision settings, but much longer
  computing times, the codes agree in the synchronous gauge up to a
  very high level that would not contribute to the present errors.}.
 
\begin{table}
  \begin{center}
  \begin{tabular}{|c|c|l|}
\hline
Name & prior range & description \\
\hline
$\anone$ & $[-1,1]$ & correlation sign between $\delta_\ub$ and $\delta_{\xe}$ \\
$\antwo$ & $[0,20]$ & overall $\xet$-dependent amplitude of $\delta_{\xe}$ \\
$\Rnone$ & $[-500 , 500 ]$ & linear bubble size (in Mpc) \\
$\Rntwo$ & $[-500 , 500 ]$ & quadratic bubble size (in Mpc)\\
$\Rnthree$ & $[-500 , 500 ]$ & cubic bubble size (in Mpc)\\
$\alpnzero$ & $[-2,100]$ & $k$-dependent weight \\
$\alpnone$ & $[0, 100]$ & $\xet$-dependent correction \\
$\gamnzero$ & $[-10, 55]$ & power-law typical value \\
$\gamnone$ & $[2, 10]$ & $\xet$-dependent correction \\
$\gamntwo$ & $[1, 10]$ & logarithmic correction \\
\hline
\end{tabular}
\caption{Free parameters modelling uncertainties on perturbed
  ionization spectra, with respective top-hat prior edges. More
  details can be found in the appendix~\ref{appendix}.}
  \label{tab:reio}
  \end{center}
\end{table}

We then turned on the blurring and lensing-blurring effects, while
adding the ten new free parameters encoding reionisation sources, (see
appendix~\ref{appendix}), modelling our ignorance on reionisation
details with conservative flat priors ranges shown in
table~\ref{tab:reio}. These prior ranges have been taken by
multiplying (and dividing) the expected value by typically two orders
of magnitude for amplitude parameters, and by one order of magnitude for
power law exponents.

\begin{figure}
  \begin{center}
    \includegraphics[width=\onefigw]{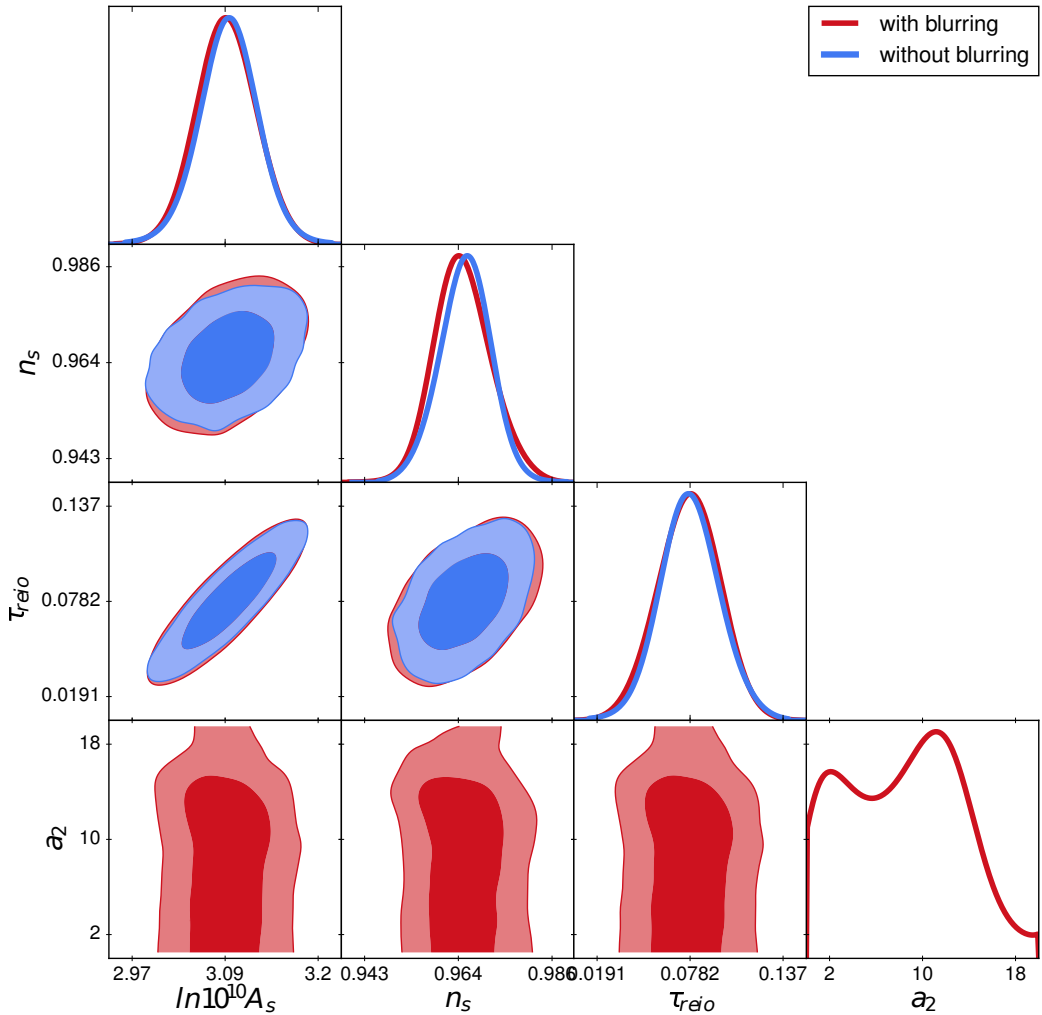}
    \caption{Posteriors and correlations (at the $68\%$ and $95\%$ levels)
      between the $\Lambda$CDM parameters, $\ln( 10^{10} A_\us)$,
      $n_\us$, $\tau_\ureio$, and the parameter $\antwo$ of the
      perturbed reionisation model.}
    \label{fig:cosmo_a2}
  \end{center}
\end{figure}
We ran again 8 Markov chains until all parameters reached a
Gelmann-Rubin convergence criterium $R-1 \leq 0.018$. Going even below
would be extremely CPU-consuming due to presence of several
reionisation parameters poorly constrained by the data. Compared to
the previous run in the Newtonian gauge, the largest variation of the
the error is for $(n_\us,\tau_\ureio)$, with respectively a
$(10\%,6\%)$ increase. All other errors vary by $5\%$ at most. The
mean value of $\tau_\ureio$ is shifted by 0.12$\sigma$, while all
other shifts are below $\pm 0.1\sigma$. We conclude that within the
accuracy of Planck and of a standard parameter extraction pipeline,
blurring has a negligible impact on the measurement of cosmological
parameters, excepted for a very small widening of the error on
$\Lambda$CDM parameters by $\sim 10\%$ at most. In two-dimensional
posteriors, we find a very small correlation between
$\{n_\us,\tau_\ureio,\ln( 10^{10} A_\us)\}$ and the reionisation
parameter $\antwo$ when $\antwo$ exceeds $\sim 10$. This parameter has
a very specific meaning: it controls the response of the
ionisation-to-baryon bias $\delta_{\xe}/\delta_\ub$ to the average
ionisation fraction $\xet$ in the large wavelength limit. However,
with our parametrisation and prior ranges, $\antwo$ plays a simple
role: it is the main parameter controlling the overall order of
magnitude of the ionisation fluctuations $\delta_{\xe}$. For
relatively large value, $\antwo \geq 10$, and thus for quite
inhomogeneous reionisation scenarios, the posterior distribution for
$n_\us$ gets a bit wider. The current fit to simulation being at
$a_2 = 9.64$, this suggests that lensing-blurring may actually
slightly affect the accuracy in the determination of the $\Lambda$CDM
parameters.

\begin{figure}
  \begin{center}
    \includegraphics[width=\onefigw]{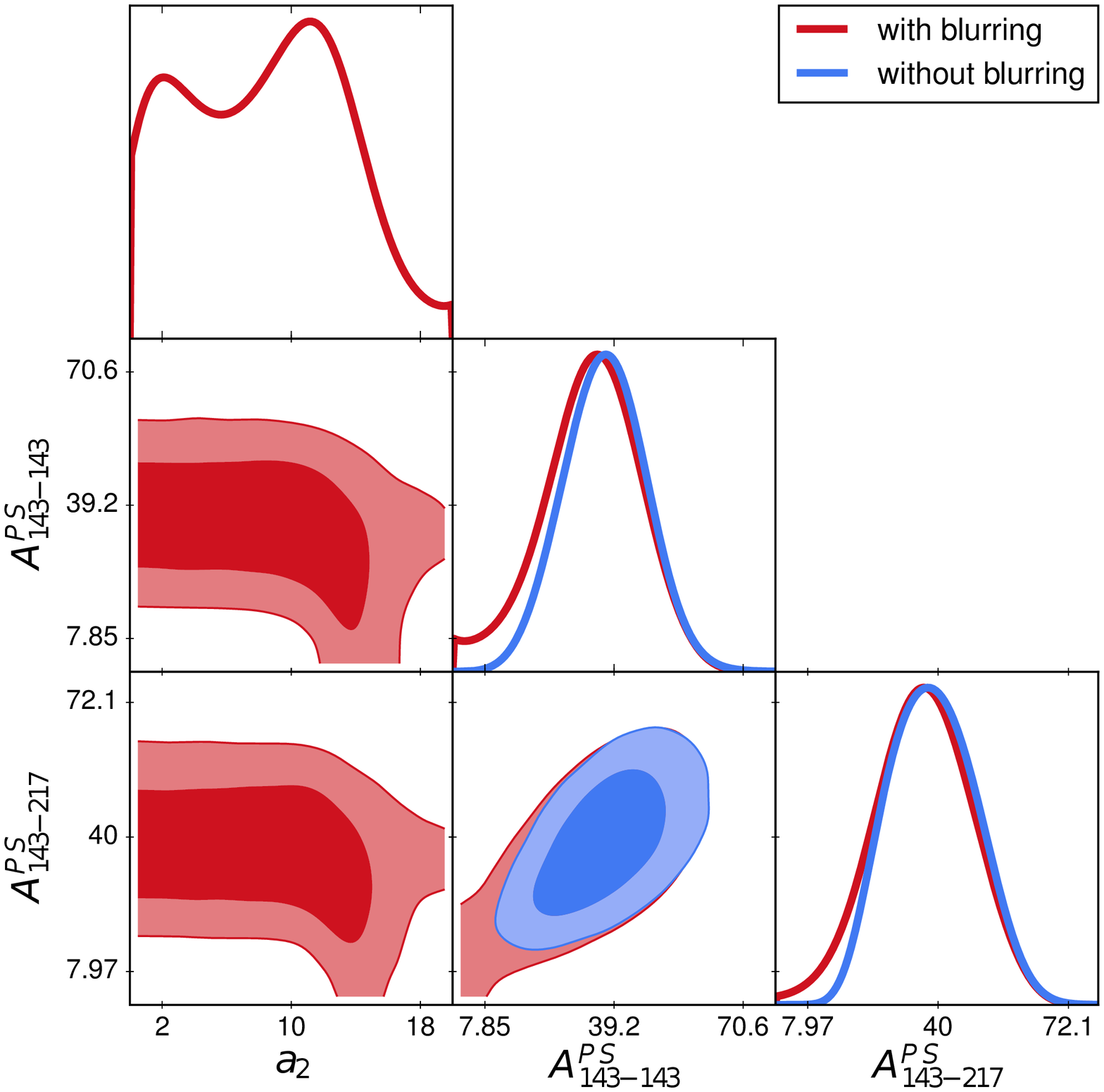}
    \caption{Posteriors and correlations (at the $68\%$ and $95\%$
      levels) between the CMB point-source foreground parameters
      $(A^\mathrm{PS}_{143-143},A^\mathrm{PS}_{143-217})$ and the
      parameter $\antwo$ of the perturbed reionisation model.}
    \label{fig:nuisance_a2}
  \end{center}
\end{figure}
We also checked that in most cases, the posterior of Planck nuisance
parameters does not change significantly when blurring effects are
switched on. Thus the absence of sensitivity of the data to blurring
is really due to the smallness of the effect, rather that to
marginalization over foregrounds and systematics. The only exception,
clearly seen in two-dimensional contours, is a degeneracy between
$\antwo$ and the nuisance parameters modelling the amplitude of
point-source foregrounds in difference frequency bands:
$A^\mathrm{PS}_{100-100}$, $A^\mathrm{PS}_{143-143}$,
$A^\mathrm{PS}_{143-217}$,
$A^\mathrm{PS}_{217-217}$~\cite{Aghanim:2015xee}. For large values
$\antwo \geq 10$, these amplitudes can be reduced. This is
particularly true at the frequency of $143\,\GHz$: we find that with
$\antwo \simeq 13$, the parameters $A^\mathrm{PS}_{143-143}$ and
$A^\mathrm{PS}_{143-217}$ are compatible with zero. This means that a
large blurring effect mimics the impact of point source contamination
on the temperature spectrum. Blurring effects can thus be relevant at
least in one case: when one tries to infer the point source
temperature spectrum from CMB observations. But again, if one assumes
the fiducial reionisation model for which $\antwo=9.64$, the impact of
such a correlation remains small.

Even though our Bayesian analysis of current data is only very weakly
sensitive to blurring, there are some interesting parameter
combinations in which the blurring effect is enhanced. This happens
for example when the blurring potential is specifically tuned such
that it is more strongly correlated with lensing. But because these
are so finely tuned models they are irrelevant in the result of a
Bayesian analysis.

\subsection{Reionisation parameters}

We investigated whether the blurring signal is sufficiently large to
constrain the parameters of perturbed reionisation.  We expect that
when reionisation is assumed to be strongly inhomogeneous, the
blurring effects will become too large to be compatible with the
data. Indeed we do find a bound on one parameter: $\antwo < 16$
($68\%$ confidence limit), assuming a flat prior on
$\antwo>0$). Compared to the fiducial value expected from reionisation
simulations, $\antwo=9.64$, such a limit confirms the weak sensitivity
of the data to the amount of inhomogeneities. Let us notice however
that both values are not orders of magnitude apart such that
competitive bounds could be in the reach of the future CMB
experiments.

The other reionisation parameters are, however, totally unconstrained
in the ranges described in Table~\ref{tab:reio}. This is not
surprising, considering that we are only probing these parameters
through small modulations of the CMB sky, and that we have introduced
ten of them.

\subsection{Investigating a possible connection with the $A_\uL$ anomaly}

The community identified a peculiarity in the Planck data. The plain
$\Lambda$CDM model does provide a very good fit to the data, with a
very satisfactory reduced $\chi^2$, $p$-value and level of
residuals. In $\Lambda$CDM, the amplitude of the CMB lensing effect in
the temperature power spectrum is fixed, because the six model
parameters define the amplitude of the matter power spectrum in the
recent universe, and thus the variance of the CMB lensing potential
and of the associated lensing deflection field.

However, as an academic exercise, one can multiply the amplitude of
the lensing potential spectrum used to compute the lensing effect in
the temperature power spectrum by a factor $A_\mathrm{L}$. Values
$A_\mathrm{L} \neq 1$ are in principle unphysical. It is however worth
noticing that when $A_\mathrm{L}$ is treated as a free parameter, the
data prefer higher than unity values, $A_\mathrm{L} = 1.22 \pm 0.10$
($68\%$CL, Planck 2015 TT+lowP)~\cite{Ade:2015xua} or $A_\mathrm{L} =
1.243 \pm 0.096$ ($68\%$CL, Planck 2018
TT+lowE)~\cite{Aghanim:2018eyx}. The level of tension with respect to
$A_\mathrm{L}=1$ does not decrease when adding high-$\ell$
polarization data. All this means that assuming an artificially large
level of smoothing of the temperature spectrum (and to a much lesser
extent, of the $E$-polarisation spectrum) by lensing allows to absorb
a non-negligible part of the residuals of the $\Lambda$CDM best
fit. Note that the parameter $A_\mathrm{L}$ should not be confused
with another parameter $A_\mathrm{L}^{\phi \phi}$ discussed
in~\cite{Ade:2015xua}. That parameter is a test of the compatibility
between the temperature spectrum and the extracted CMB lensing
spectrum. That test is independent of the previous one, and returns no
anomalous behaviour.

It is not obvious that there is something to learn from this
$A_\mathrm{L}$ test. A priori, we could transform each one of the few
hundreds of coefficients appearing in the equations defining linear
cosmological perturbation theory into a free parameter. We would then
expect that $\sim 1.3\%$ of them have a best-fit value $2.5 \sigma$
away from their theoretical value. $A_\mathrm{L}$ could just be within
these $1.3\%$. On the other hand, it is always worth considering
whether there could be a convincing physical explanation mimicking
this effect and allowing to increase the likelihood of the best fit
model.

The residuals of the best-fit $\Lambda$CDM model to the Planck data
show some vaguely oscillatory features roughly in phase with lensing
effects the range $1100<\ell<2000$ (see Fig.~24 in
Ref.~\cite{Aghanim:2018eyx}). Since the blurring-lensing correlation
effect also has oscillatory patterns, it is worth checking whether the
observed ``anomaly'' could actually arise from neglecting this
effect. Our strategy is to run simultaneously with blurring effects
turned on and a free parameter $A_\mathrm{L}$. If the blurring effects
could fit some of the residuals giving rise to the ``$A_\mathrm{L}$
anomaly'', the posterior for $A_\mathrm{L}$ would become more
compatible with $A_\mathrm{L}=1$ and the ``anomaly'' would be
explained.

Our run with $17$ model parameters ($6$ $\Lambda$CDM parameters, $10$
perturbed reionisation parameters and $A_\mathrm{L}$) plus Planck
nuisance parameters gives some marginalized bounds
$A_\mathrm{L}=1.23^{+0.09}_{-0.11}$ ($68\%$CL, Planck 2015 TT+lowP),
to be compared with $A_\mathrm{L} = 1.22 \pm 0.10$ in absence of
blurring. Once again, we find that the impact of blurring is
negligible. We also checked the absence of correlation in the
two-dimensional contour plots between $A_\mathrm{L}$ and the
parameters describing perturbed reionisation. The reason is the
$\pi/2$ phase shift between the oscillations in the lensing-lensing
and blurring-lensing effects. This shift, that makes it impossible to
reproduce a lensing-like signal in the data from blurring, appears to
be a generic feature, present even when the parameters of perturbed
reionisation are varied.

\section{Conclusions}
\label{sec:con}

We have computed, for the first time, the inclusive impact of the
reionisation blurring and its correlation with weak gravitational
lensing. The blurring of the cosmic microwave background is induced
from ionised gas at the time of reionisation, plus other velocity- and
GR-induced effects that we have considered, and is a prediction in the
standard cosmological model. In principle it offers a probe into the
epoch of reionisation allowing to study not only the optical depth of
reionisation, but also the dynamics of perturbations during
reionisation.

However, we find that the signal is small and that a reasonable
parameter space for reionisation cannot be constrained with current
data. Since Planck is already cosmic variance limited in the lower and
medium temperature multipoles, this is unlikely to change in the near
future.

We have studied in particular the correlation between lensing and
blurring that in principle could have been observable and explain the
measured lensing anomaly. However we find that the correlation between
the lensing and blurring potential is weak on the large scales that
are particularly important in the lensing computation. As a result the
overall correlation is small. In addition we find a generic phase
shift between lensing and the lensing-blurring correlation that makes
it impossible to mistake a blurring signal for enhanced lensing. Our
analysis shows that blurring cannot explain the lensing anomaly and
that $A_\uL$ remains unaffected within the statistical errors.

Since reionisation blurring is a non-speculative effect, it should be
included in the data analysis to obtain unbiased parameters. However,
we find that the cosmological parameters are not affected much by the
blurring signal. In the case of a larger than expected reionisation
parameter $\antwo>10$, the most affected parameter is $n_\us$, whose
bounds get broadened by $10\%$. In that case, blurring effects can be
partially degenerate with foregrounds from point sources. However, for
$\antwo<10$, the blurring impact remains negligible. In the future,
reionisation models could be better inferred from a combination of
observations and N-body/reionisation simulations, such that all
reionisation parameters including $\antwo$ could be fixed to their
expected value.

Overall we conclude that for the current experiments blurring does not
play an important role in the CMB data analysis based on the angular
power spectra. However it is much bigger than any other non-linear
correction along the line-of-sight, beyond lensing, causing
corrections of a few percent compared to lensing. This is at the limit
of the PLANCK data sensitivity. With a more precise experiment,
reionisation blurring may be observable and this could open a new
window into the epoch of reionisation.

It should be noted that our analysis does not include several other
secondary effects. While we consider photons scattered out of the
line-of-sight, we are not including photons that are emitted into our
line-of-sight. This contribution was discussed in our previous paper,
see Ref.~\cite{Fidler:2017irr}, and is structurally very different
from the blurring signal. It dominates the small scales and acts
similar to the Ostriker-Vishniak effect~\cite{Ostriker:1986,
  Vishniac:1987, Jaffe:1997ye}. Both non-linear effects originating
from reionisation are of roughly comparable order of magnitude, but
only blurring is correlated to lensing, providing a boost in
amplitude.

Finally, a word of caution about our analysis is in order. We have
been focused on the impact of blurring onto the power spectra
only. There is still the possibility that blurring impacts various
$N$-point functions of the temperature and polarisation anisotropies,
in particular any blurring imprints onto the trispectrum could, in
principle, bias the lensing extraction
pipeline~\cite{Namikawa:2017uke}. We leave however such a study for a
future work.

\appendix

\section{Reionisation parametrisation}
\label{appendix}

\subsection{Ionising power spectra}

The power spectra for the ionized fraction and its correlation to
baryons are given in equations~\eqref{eq:reio_fit} and
\eqref{eq:reio_cross}. The functions entering these equations have
been fitted against the ionised background fraction $\xet$, from the
tabulated points obtained in the reionisation history model considered
in Ref.~\cite{Mao:2008ug} (see figure~2 and table~3), which is based
on the numerical simulations of Ref.~\cite{McQuinn:2007dy}. The shape
of these functions is motivated by their good fit to the tabulated
data, and, their robustness with respect to
extrapolation~\cite{Clesse:2012th, Fidler:2017irr}. For instance, the
amplitude functions $\Ai$ (and $\Ax$) are modelled as exponential functions to
ensure their positivity whereas a simple Taylor expansion has been
chosen for the typical radius $\Ri$ (and $\Rx$) of ionised
structures. They read
\begin{equation}
\begin{aligned}
\Ai^2(x) &= \aione e^{\aitwo x}, & \quad \Ax(x) &= \axone e^{\axtwo x}, \\
\Ri(x) &= \Rione x + \Ritwo x^2 + \Rithree x^3, & \quad \Rx(x) & =
\dfrac{\Rxzero}{1 + e^{\Rxone(x-\Rxtwo)}} \,,\\
\alphai(x) &= \max \left(-2,\alpizero + \alpione  x \right), & \quad
\alphax(x) & = \alpxzero + \alpxone x e^{\alpxtwo x}, \\
\gammai(x) & = \gamizero x^{\gamione} \left[-\ln(x)\right]^{\gamitwo}.
& \quad \phantom{\dfrac{1}{2}}
\end{aligned}
\label{eq:functionaldep}
\end{equation}
The fiducial values reproducing the results of Refs.~\cite{Mao:2008ug,
  McQuinn:2007dy} are given in Table~\ref{tab:reioparams} and the
resulting ionising power spectra have been plotted in
figure~\ref{fig:reiopk}.

\begin{table}
  \begin{center}
  \begin{tabular}{|c|c|c|c|c|c|c|}
\hline
Function & Parameter & Value & Parameter & Value & Parameter & Value \\
\hline
$\Ai(x)$ & $\aione$ & $0.09$ & $\aitwo$ & $9.64$ & & \\
\hline
$\Ri(x)$ & $\Rione$ & $17.81\,\Mpc$ & $\Ritwo$ & $-58.03\,\Mpc$ & $\Rithree$ &
$55.29\,\Mpc$ \\
\hline
$\alphai(x)$ & $\alpizero$ & $-2.93$ & $\alpione$ & $10.04$ & & \\
\hline
$\gammai(x)$ & $\gamizero$ & $16.13$ & $\gamione$ & $2.20$ &
$\gamitwo$ & $1.23$ \\
\hline
\hline
$\Ax(x)$ & $\axone$ & $0.306$ & $\axtwo$ & $4.727$ & & \\
\hline
$\Rx(x)$ & $\Rxzero$ & $0.59\,\Mpc$ & $\Rxone$ & $26.24$ & $\Rxtwo$ & $0.66$
\\
\hline
$\alphax(x)$ & $\alpxzero$ & $-0.10$ & $\alpxtwo$ & $10.73$ & & \\
\hline
\end{tabular}
\caption{Best fit values for the parameters entering into the
  functional dependency of the ionising power spectra of
  equations~\eqref{eq:reio_fit} and \eqref{eq:reio_cross} with respect to
  the ionised fraction $x$. The functions $A(x)$ encode their
  amplitude, $R(x)$ the typical size of ionised structures while the
  other functions represent different power-law exponents.}
\label{tab:reioparams}
  \end{center}
\end{table}

\subsection{Simpler model for CMB marginalisation}

For the CMB data analysis in presence of blurring and
blurring-lensing, because we are only interested in marginalising over
the ionising sources, as explained in section~\ref{sec:cmb}, we have
chosen a much simplified approach in which we postulate that the
ionised fraction itself behaves as:
\begin{equation}
  \delta_{\xe}(\eta,k) = \An(\xet) \left(1 -
  \xet \right) \left\{1 + \alphan(\xet) k \Ri(\xet) + \left[k
    \Rn(\xet) \right]^2 \right\}^{-\gamman(\xet)/4}
  \delta_{\ub}(\eta,k)\,.
\label{eq:deltaxeeff}
\end{equation}
This form is simply the same as the square root of the power spectrum
$\xet^2 P_{\delta_{\xe}\delta_{\xe}}$, with all the parametrising
functions $\An$, $\Rn$, $\alphan$ and $\gamman$ assuming the same
functional dependency with respect to $\xet$ as in
equation~\eqref{eq:functionaldep}, namely
\begin{align}
  \label{eq:An}
  \An(x) & = \sign{(\anone)} \sqrt{|\anone|} e^{\frac{\antwo x}{2}},\\
\label{eq:Rn}
  \Rn(x) & = \Rnone x + \Rntwo x^2 + \Rnthree x^3,\\
\label{eq:alpn}
  \alphan(x) & = \max(-2,\alpnzero + \alpnone x),\\
\label{eq:gamn}
  \gamman(x) & = \gamnzero x^{\gamnone}\left[-\ln(x) \right]^{\gamntwo}.
\end{align}
However, we do no longer fix their parameters to the realistic values
given in Table~\ref{tab:reioparams} but we rather vary them over a
wide prior range given in Table~\ref{tab:reio}.  Strictly speaking,
equation~\eqref{eq:deltaxeeff} cannot be straightforwardly used for the
computation of the cross power spectrum with baryons, $P_{\delta_\ub
  \delta_{\xe}}$. However, as can be checked in
figure~\ref{fig:reiopk}, for the best fit parameters, both spectra are
of very similar shape on scales larger than typically
$k=0.2\,\Mpc^{-1}$, which are the ones relevant for our CMB
analysis. Therefore, for the purpose of CMB marginalisation over
nuisances, equation~\eqref{eq:deltaxeeff} is sufficient. Notice
however that the sign of $\anone$ in equation~\eqref{eq:An} does not
play a role for $P_{\delta_{\xe}\delta_{\xe}}$ but does encode the
correlation sign between $\delta_{\ub}$ and $\delta_{\xe}$.

\acknowledgments

Simulations were performed with computing resources granted by
JARA-HPC from RWTH Aachen University under project jara0184. The work
of C.\ R. is supported by the ``Fonds de la Recherche Scientifique -
FNRS'' under Grant $\mathrm{N^{\circ}T}.0198.19$.

\bibliographystyle{JHEP}
\bibliography{references}

\end{document}